# MODEL FOR HIGH TEMPERATURE PHASE OF $C_{70}$ SOLID

**Sarbpreet Singh, K. Dharamvir and V.K. Jindal[1]**
Department of Physics, Panjab University, Chandigarh-160014, India

**ABSTRACT**

*Depending on the temperature, the $C_{70}$ solid crystallizes in several structures. At high temperature (T > 340K), the ellipsoidal $C_{70}$ molecule rotates freely in all directions and may be treated as a uniform thick spherical shell with inner and outer radii as the minimum and the maximum distance of C-atom from the center of the molecule. At lower temperatures the free rotations of molecules freeze out. We have calculated the lattice parameters, energies and bulk modulus at the minimum energy configuration of fcc and hcp phase of pure $C_{70}$ solid at high temperature using a simple model based on atom-atom potential.*

**INTRODUCTION**

Carbon is a remarkable elememt showing a variety of stable forms. Fullerenes are often referred as third form of carbon, others being diamond and graphite. The word "fullerene" is used to denote the whole class of closed cage molecules consisting of only carbon atoms. The most abundantly known fullerene consists of 60 carbon atoms and the solid popularly known as $C_{60}$ solid. The $C_{60}$ solid has been extensively studied both theoretically and experimentally. Another fullerene solid consisting of molecules comparising of 70 atoms of carbon, known as $C_{70}$ solid is also found in good abundance. As the molecular weight increases beyond $C_{70}$, the abundance decreases dramatically.

Whereas $C_{60}$ molecule is nearly spherical in shape with all the carbon atoms located at the vertices of a truncated icosahedron having 60 vertices, the molecule of $C_{70}$ is more like a rugby ball. The $C_{70}$ molecule can be envisioned by adding a ring of 10 carbon atoms or, equivalently, adding a belt of five hexagons around the equatorial plane of $C_{60}$ molecule which is normal to one of the five fold axes and rotating the two hemisphere of $C_{60}$ so that they fit continuously onto the belt hexagons [1] as shown in fig.1.

---

[1] Author with whom correspondence be made, *jindal@panjabuniv.chd.nic.in*

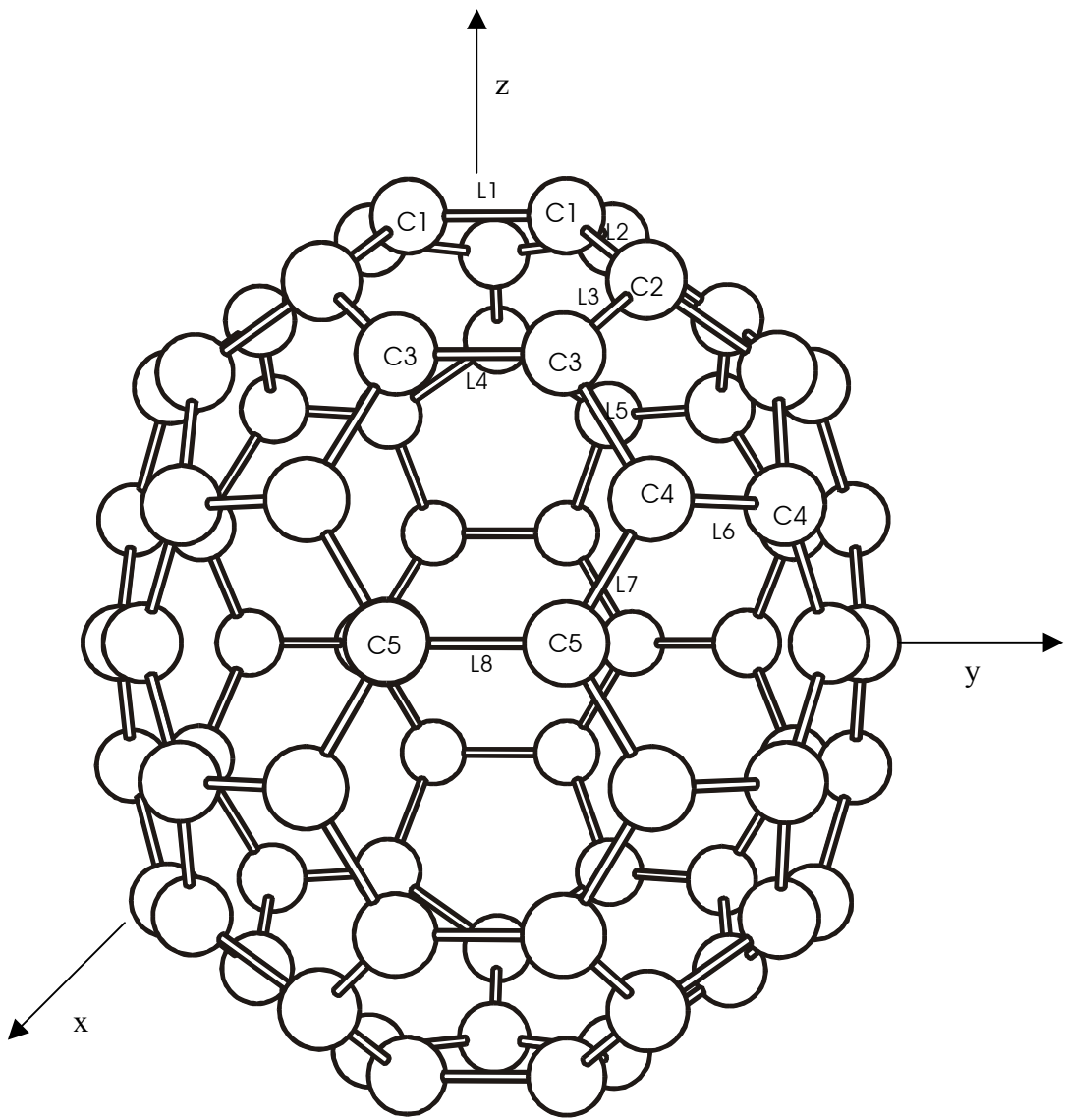

**Fig.1** The $C_{70}$ molecule with circles representing the C-atoms

For $C_{60}$ the stacking sequence is ABCABC…, at all temperatures, thus the crystal structure are face-centered cubic (fcc) and simple cubic (sc). Above room temperature, the $C_{60}$ molecules are spinning rapidly about their lattice positions and there is no orientational order and the solid crystallizes into a fcc phase. As the temperature is lowered below room temperature, the free rotations freeze and the system undergoes a phase transition from fcc to sc. For $C_{70}$, both ABCABC…. and ABAB…. sequences are observed and hence the crystal structures are fcc or hcp depending on the temperature range.

Several experimental observations have been made to study the structure[2,3] and phase transitions[1] in $C_{70}$. Bulk and the thermodynamical properties of $C_{70}$ solid have also been experimentally measured[4]. Theoretical models for fullerene solids are generally based on model potentials for C-C interaction which are then summed over to obtain molecule-molecule potentials. Generally, the fullerene molecules are considered to be stiff and rigid while in the solid. Although some calculations[5] have been done based on molecular-dynamics calculations(MD) to study the structural properties of $C_{70}$ using Lennard-Jones potential, an exercise needs to be attempted for the detailed temperature and pressure effects and to assess the usefulness of the simple model chosen here. A simple model based approach would be very useful for calculations of various bulk, lattice and thermodynamic properties and thus provides a quick comparison with the measured results.

**PHASE TRANSITIONS IN $C_{70}$**

Depending on the temperature, $C_{70}$ solid has been found to crystallize in various crystal phases[6]. At high temperatures (T > 340K), solid $C_{70}$ shows no long-range orientational order of its molecules. In this temperature regime, the fcc phase with freely rotating molecules is the most stable phase with ideal hexagonal closed packed phase almost equally stable. The molecular centers are arranged on fcc lattice with four $C_{70}$ molecules per unit cell and on hcp lattice with two molecules per unit cell. As the temperature is lowered below $T_{01}$(=340K), a phase transition to a phase with long-range orientational order of the molecules takes place. In this temperature range, the $C_{70}$ molecules rotate only about the five fold axis. In this range, the long-range orientational order breaks the cubic symmetry and the crystal distorts into a rhombohedral lattice. As the temperature is further lowered below $T_{02}$(=280K), the free rotation about the long axis also becomes frozen and the crystal phase below this temperature is a monoclinic structure. In the intermediate temperatures the $C_{70}$ molecule may, therefore, be modeled as an ellipsoid of revolution which on further increase in temperature above $T_{01}$ can be modeled as a spherical shell. Various possible structures for $C_{70}$ at different temperature ranges have been summarized in fig.2.

**Fig.2** Various structures of $C_{70}$ in different temperature ranges[6].

| Low temperature phase (T<280 K) | Intermediate temperature phase (280 K<T<340 K) | High temperature phase (T>340 K) |
|---|---|---|
| **Monoclinic** Molecular orientations ordered | **Rhombohedral** Molecules rotate about long axis. | **fcc and hcp** Disordered molecular orientations |

**THEORETICAL MODEL AND CALCULATIONS**

**General Interaction Model**

The knowledge of the interaction potential is the basic ingredient to calculate the bulk, structural and dynamical properties of any system. Following our earlier success in modeling C-cluster molecular solids, we assume that atom-atom potential govern the inter-molecular potential in $C_{70}$ solids also. The intermolecular potential energy $U_{l\kappa,l'\kappa'}$ between the two $C_{70}$ molecules identified by κ molecule in unit cell l, κ' molecule in cell l', can be written[7] as

$$U_{l\kappa,l'\kappa'} = \sum_{ij} V(r_{ij}) \tag{1}$$

where $r_{ij}$ is the distance between C-atom of molecules lκ and l'κ'. In case of $C_{70}$ the summation includes all the 70 atoms in each of the molecules and V(r) is the C-C potential for the two C atoms situated at r distance apart, and is of the form

$$V(r) = -\frac{A}{r^6} + B.\exp(-\alpha r) \tag{2}$$

The interaction parameters A, B and α are taken from the set provided by Kitaigorodski, which have already been used successfully[7].

From the knowledge of the crystal structure which provides the lattice positions, the total potential energy Φ can be obtained numerically by carrying out the lattice sums. In order to obtain the total potential energy Φ, we need to sum $U_{l\kappa,l'\kappa'}$ (Eq. (1)) for all pair-wise molecular interactions. For this summation, we need all the atom-atom distances between various molecules.

$$\Phi = \frac{1}{2}\sum_{l\kappa,l'\kappa'} U_{l\kappa,l'\kappa'} \tag{3}$$

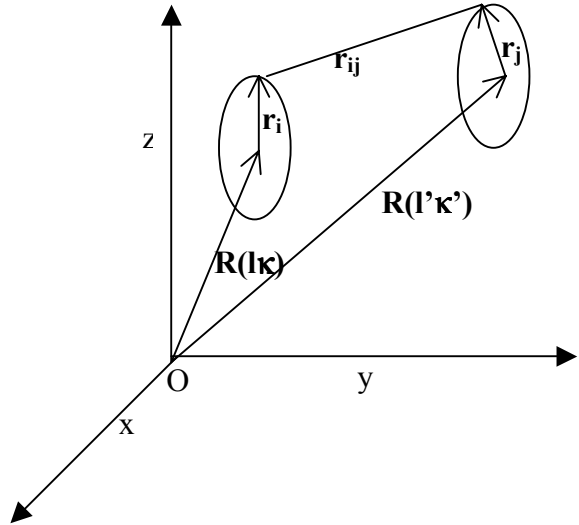

**Fig.3** Diagram representing the positions of $i^{th}$ and $j^{th}$ atoms on two different $C_{70}$ molecules.

Knowing the position vectors of the centers of the molecules as **R(l**$\kappa$**)** and **R(l'**$\kappa'$**)**, and atomic vectors **r**$_i$ and **r**$_j$, the atom-atom distance $r_{ij}(l\kappa,l'\kappa')$, as shown in fig.3, is written as

$$r_{ij}(l\kappa,l'\kappa') = | \mathbf{R}(l'\kappa') + \mathbf{r_j} - \mathbf{R}(l\kappa) - \mathbf{r_i} | \tag{4}$$

Therefore, knowing the crystal structure ($R(l\kappa)$) and knowing the atomic coordinates of $C_{70}$ from the molecular geometry, the atom-atom distances can be obtained which finally leads to the total potential energy (Eq. (3)).

In $C_{70}$ molecule there are five inequivalent carbon sites (C1, C2,….C5) and eight distinct bond lengths (L1, L2, ……L8) as shown in fig.1. These eight bond lengths found experimentally by neutron inelastic scattering (NIS) [8] are listed Table 1.

**Table 1** Experimental bond lengths of $C_{70}$ (in A)

| Bond | $C_1$-$C_1$ | $C_1$-$C_2$ | $C_2$-$C_3$ | $C_3$-$C_3$ | $C_3$-$C_4$ | $C_4$-$C_4$ | $C_4$-$C_5$ | $C_5$-$C_5$ |
|---|---|---|---|---|---|---|---|---|
| NIS | 1.460 | 1.382 | 1.449 | 1.396 | 1.464 | 1.420 | 1.415 | 1.477 |

Various groups have also calculated these lengths using tight binding molecular dynamics (TBMO)[9], the local density approximation (LDA)[10], modified neglect of differential overlap(MNDO)[11] and Hartree-Fock(HA) methods and got the bond lengths ranging from 1.356A to 1.475A.

We have generated the atomic coordinates of $C_{70}$ by adding a belt of five hexagons around the equatorial plane of $C_{60}$ molecule which is normal to one of the five fold axes and rotating the two hemisphere of $C_{60}$ so that they fit continuously onto the belt hexagons and considering only two types of bond lengths with single and double bond lengths as 1.40A and 1.45A. From the coordinates so obtained it has been observed that the distance of the nearest and the farthest carbon atoms from the center of the $C_{70}$ molecule taken as the semi minor and the semi major axis are a = 3.498A and b = 4.112A respectively.

**Free rotation model**

For the case of freely rotating molecules at high temperatures above 340K, as in our case, the orientational correlation vanishes and the molecules rotate freely in all directions. We model the $C_{70}$ molecule by assuming all the 70 C-atoms to be distributed uniformly within a thick spherical shell of inner and outer radii as the semi minor and the semi major axis of the $C_{70}$ molecule, with density ρ as shown in fig.4.

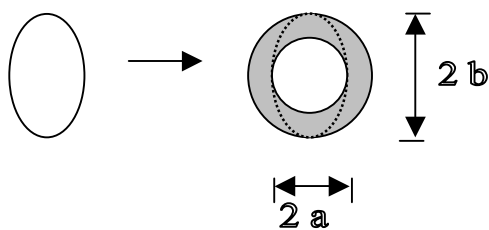

**Fig.4** High T equivalent spherical thick shell $C_{70}$ molecule

The density can be calculated as $\rho = 70/\Delta V$, where $\Delta V$ is the volume of the thick shell enclosing the 70 carbon atoms and is given by $\Delta V = \frac{4}{3}\pi(b^3 - a^3)$. In this case of free rotation of molecules, the summation over all the 70 atoms in each of the molecules can be replaced by the integration over the spherical shells. The interaction between the two $C_{70}$ shells can be calculated by considering a small volume element $dV_1$ on one shell with position vector $\mathbf{r_1}$ and $dV_2$ on another shell with position vector $\mathbf{r_2}$ (Fig.3) and can be written as

$$V(R,a,b) = \int\int (\rho dV_1)(\rho dV_2)\left[-\frac{A}{|r_2-r_1|^6} + B\exp(-\alpha|r_2-r_1|)\right] \quad (5)$$

In this way the expression for V has been analytically obtained which is a function of R, the distance of separation of the centers of the two $C_{70}$ thick spherical shells and their geometrical parameters, a and b. The evaluation of the final expression has been described in the appendix.

The summation over various lattice points is obtained for intermolecular distances up to some distance $R_L$. The potential energy as a function of $R_L$ has been plotted in fig.5 which shows good convergence ($\sim 7\times10^{-5}$ %) when the summation is restricted to the distances upto 85 A. In our calculations the summation up to a intermolecular distances of 90 A has been done.

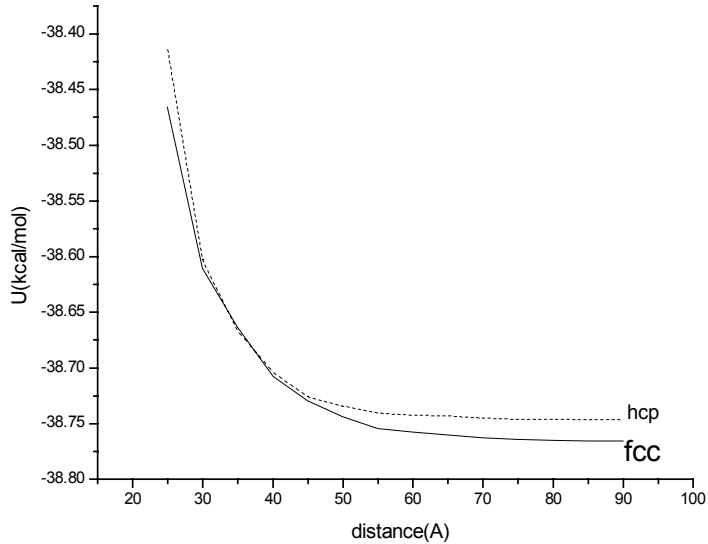

**Fig.5** Potential energy as a function of the distance of summation.

The potential energy thus obtained as a function of lattice parameter has been plotted in fig.6. The lattice parameters (a) corresponding to the minimum potential energy and the minimum energy ( Φ ) for the equilibrium configuration for both the fcc and hcp systems are presented in Table 2. In this table we also present the results from the molecular-dynamics (MD) calculations done by Pickholz *et. al.*[5] and the experimental results for comparison. Here in our calculations the hcp structure has been taken to be ideal with c/a ratio as 1.63.

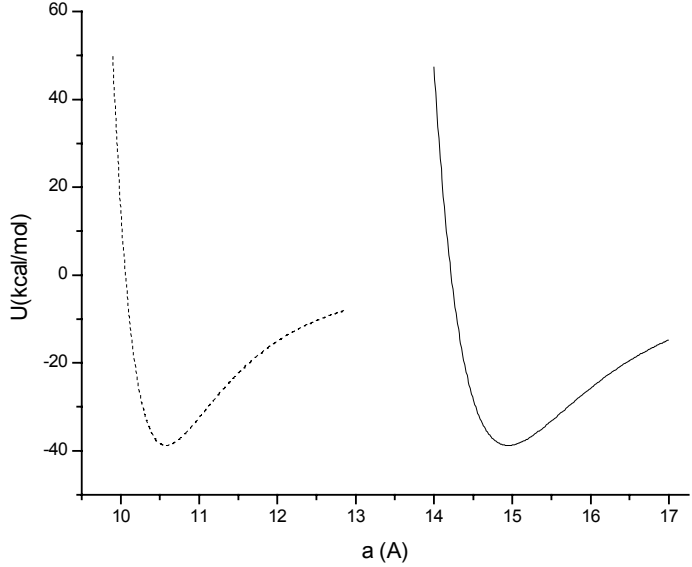

**Fig.6** Calculated potential energy of the $C_{70}$ solid as a function of lattice parameter for fcc and hcp structures at high temperatures. The broken line is hcp and continuous line is fcc result.

**Table.2** Calculated and experimental structure and energies for $C_{70}$ lattice

| Structure | Lattice parameters (A) | | | Potential energy (kcal/mol) | | |
|---|---|---|---|---|---|---|
| | Our calc. | Exp.[2-3] | Other calc.[5] | Our calc. | Exp.[12] (at 739K) | Other calc.[5] (at 739K) |
| Fcc | a =14.96 | 15.01 | 14.80(at room Temp.) | 38.76 | 43.06 | 44.74 |
| Hcp | a =10.58 c = 17.25 | 10.56 17.23 | 10.20(at 200K) | 38.74 | 43.02 | 44.70 |

We compare the numerical results for the various physical quantities obtained for pure $C_{70}$ solid at high temperature phase using a simplified model based on atom-atom potentials. Comparison of the calculated lattice parameters (a) and potential energies (Φ) with the experimental values and other calculations done by Pickholz et. al. [5], reveals that the present calculation is able to reproduce the structure and cohesive energies quite well for both the fcc and hcp structures. Comparing the calculated potential energies for fcc and hcp structures it is found that fcc phase has energy lower than hcp phase by only 0.05%.

**Bulk Modulus**

The potential energy varies with the application of the hydrostatic pressure 'p' such that

$$\Phi_p = \Phi + p\Delta V \qquad (6)$$

where $\Phi_p$ is the total potential energy of the system at pressure p and $\Delta V$ is the change in the volume of unit cell per molecule of $C_{70}$ because of the application of the pressure. Therefore, bulk modulus B ($= -V\partial p / \partial V$) can be calculated at minima by knowing the p-V curve for the minimized new potential energy. The plot of volume of unit cell per molecule as a function of pressure for both the systems (fcc and hcp) are presented in Fig.7. The calculated bulk modulus for both fcc and hcp structures are presented in Table 3. The volume of unit cells per molecule for fcc and hcp structures at minimum energy configuration at p=0 are also tabulated in Table 3.

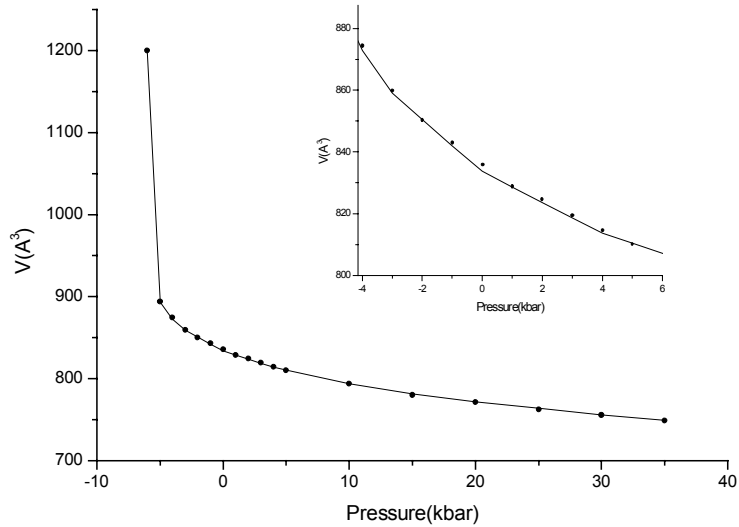

**Fig.7** Calculated volume per molecule of the $C_{70}$ as a function of pressure for fcc and hcp structures at high temperatures. The dots represent hcp and continuous line is fcc result.

**Table.3** Calculated bulk modulus for $C_{70}$ lattice at high temperatures.

| Structure | Bulk Modulus (GPa) | | Unit cell volume per molecule ($A^3$) |
|---|---|---|---|
| | p=0 | p=30 kbar | |
| Fcc | 13.26 | 104.9 | 835.3 |
| Hcp | 13.02 | 103.2 | 835.9 |

**SUMMARY AND DISCUSSION**

We calculated the numerical values for the various physical quantities like lattice parameters (a), potential energies ($\Phi$) and the bulk modulus for pure $C_{70}$ solid at high temperature phase using a simplified model based on atom-atom potentials and compared them with some of the known experimental values and other calculations. For $C_{70}$ solid at high temperatures, fcc phase has energy lower than hcp phase by only 0.05%. Thus we can conclude that fcc structure is slightly more stable than ideal hcp structure. Though no experimental data was available for comparison of the bulk modulus, however the bulk modulus on comparison with $C_{60}$ solid [ 7 ] shows that the $C_{70}$ solid is more compressible.

Further in order to calculate the lattice dynamics, internal vibrational modes would need to be included. Also for intermediate and low temperature, the C-atoms can no longer be assumed to be distributed uniformly over a thick shell. For intermediate temperatures, all C-atoms are distributed over the surface of $C_{70}$ molecule and for low temperatures, where all the rotations freeze out, the total interaction can be found by pair wise sum of all the 70 C-atom-atom potentials on all the molecules. The work in this direction is in progress and will be reported.

**Acknowledgements:**

The authors gratefully acknowledge the financial support for the research project "Phonon Dynamics of Fullerenes and Derivatives", SP/S2/M13/96 from the Department of Science and Technology, Government of India, New Delhi, India under which the present work was done.

## APPENDIX

As discussed earlier at higher temperature rugby ball shaped $C_{70}$ molecule can be replaced by a thick spherical shell and the interaction potential can be obtained by integrating over the two shells instead of atom-atom summation. As already discussed for this case we can write the interaction potential by considering the two volume elements $dV_1$ and $dV_2$ with position vectors $\mathbf{r_1}$ and $\mathbf{r_2}$ on two shells as given by equation (5), where $\rho$ is the density of carbon atoms within the thick shell volume and $dV_1$ and $dV_2$ are taken as the volumes of the thick ring along the shell surfaces and are given by

$$dV_1 = (2\pi r_1 \sin\theta_1).r_1 d\theta_1.dr_1 \qquad \text{A1}$$

$$\text{and} \quad dV_2 = 2\pi(r_2 - r)\sin\theta_2.(r_2 - r)d\theta_2.dr_2 \qquad \text{A2}$$

where R is the distance between the centers of the two shells. Substituting these values in equation (5) and integrating over the two volumes the final equation for the total interaction between the two shells can be obtained very easily and is represented by

$$V(r,a,b) = U_{att} + U_{rep} \qquad \text{A3}$$

where $U_{att}$ and $U_{rep}$ are the attractive and the repulsive contributions and are given by

$$U_{att} = \pi^2 \rho^2 A U'_{att}$$

where $U'_{att} =$

$$\frac{b}{2R}\log\left|\frac{(R-2b)(R+a+b)(R+a-b)}{(R+2b)(R-a+b)(R-a-b)}\right| + \frac{1}{2}\log\left|\frac{R^2(R+a+b)(R-a-b)}{(R+2b)(R-2b)(R-a+b)(R+a-b)}\right|$$

$$+\frac{a}{2R}\log\left|\frac{(R-2a)(R+a+b)(R-a+b)}{(R+2a)(R-a-b)(R+a-b)}\right| + \frac{1}{2}\log\left|\frac{R^2(R+a+b)(R-a-b)}{(R+2a)(R-2a)(R+a-b)(R-a+b)}\right|$$

$$+\frac{1}{6R}\left[(3b+2R)\log\left|\frac{(R+2b)(R-a+b)}{R(R+a+b)}\right| + (b^2+bR)\left(\frac{1}{R+2b} - \frac{1}{R} - \frac{1}{R+a+b} + \frac{1}{R-a+b}\right)\right]$$

$$-\frac{1}{6R}\left[(3a+2R)\log\left|\frac{R(R+a+b)}{(R+2a)(R+a-b)}\right| + (a^2+aR)\left(\frac{1}{R+a+b} - \frac{1}{R+2a} - \frac{1}{R+a-b} + \frac{1}{R}\right)\right]$$

$$+\frac{1}{6R}\left[(3b-2R)\log\left|\frac{R(R-a-b)}{(R+a-b)(R-2b)}\right| + (-b^2+bR)\left(\frac{1}{R} - \frac{1}{R+a-b} - \frac{1}{R-2b} + \frac{1}{R-a-b}\right)\right]$$

$$-\frac{1}{6R}\left[(3a-2R)\log\left|\frac{(R+b-a)(R-2a)}{R(R-a-b)}\right| + (-a^2+aR)\left(\frac{1}{R-a+b} - \frac{1}{R} - \frac{1}{R-a-b} + \frac{1}{R-2a}\right)\right]$$

$$\text{A4}$$

$$U_{rep} = -\frac{4\pi^2\rho^2 B}{\alpha^2}\cdot U'_{rep}$$

where $U'_{rep} =$

$$\left(\frac{b\alpha+1}{\alpha^5 R}\right)\begin{bmatrix}-(b^2\alpha^2+2b\alpha+2)e^{-\alpha(R+2b)}+(a^2\alpha^2+2a\alpha+2)e^{-\alpha(R+a+b)}-(\alpha R+2)(b\alpha+1)e^{-\alpha(R+2b)}\\+(\alpha R+2)(a\alpha+1)e^{-\alpha(R+a+b)}+(b^2\alpha^2-2b\alpha+2)e^{-\alpha R}-(a^2\alpha^2-2a\alpha+2)e^{-\alpha(R-a+b)}\\-(\alpha R+2)(b\alpha-1)e^{-\alpha R}+(\alpha R+2)(a\alpha-1)e^{-\alpha(R-a+b)}\end{bmatrix}$$

$$-\left(\frac{a\alpha+1}{\alpha^5 R}\right)\begin{bmatrix}-(b^2\alpha^2+2b\alpha+2)e^{-\alpha(R+a+b)}+(a^2\alpha^2+2a\alpha+2)e^{-\alpha(R+2a)}-(\alpha R+2)(b\alpha+1)e^{-\alpha(R+a+b)}\\+(\alpha R+2)(a\alpha+1)e^{-\alpha(R+2a)}+(b^2\alpha^2-2b\alpha+2)e^{-\alpha(R+a-b)}-(a^2\alpha^2-2a\alpha+2)e^{-\alpha R}\\-(\alpha R+2)(b\alpha-1)e^{-\alpha(R+a-b)}+(\alpha R+2)(a\alpha-1)e^{-\alpha R}\end{bmatrix}$$

$$+\left(\frac{b\alpha-1}{\alpha^5 R}\right)\begin{bmatrix}-(b^2\alpha^2+2b\alpha+2)e^{-\alpha R}+(a^2\alpha^2+2a\alpha+2)e^{-\alpha(R+a-b)}-(\alpha R+2)(b\alpha+1)e^{-\alpha R}\\+(\alpha R+2)(a\alpha+1)e^{-\alpha(R+a-b)}+(b^2\alpha^2-2b\alpha+2)e^{-\alpha(R-2b)}-(a^2\alpha^2-2a\alpha+2)e^{-\alpha(R-a-b)}\\-(\alpha R+2)(b\alpha-1)e^{-\alpha(R-2b)}+(\alpha R+2)(a\alpha-1)e^{-\alpha(R-a-b)}\end{bmatrix}$$

$$-\left(\frac{a\alpha-1}{\alpha^5 R}\right)\begin{bmatrix}-(b^2\alpha^2+2b\alpha+2)e^{-\alpha(R-a+b)}+(a^2\alpha^2+2a\alpha+2)e^{-\alpha R}-(\alpha R+2)(b\alpha+1)e^{-\alpha(R-a+b)}\\+(\alpha R+2)(a\alpha+1)e^{-\alpha R}+(b^2\alpha^2-2b\alpha+2)e^{-\alpha(R-a-b)}-(a^2\alpha^2-2a\alpha+2)e^{-\alpha(R-2a)}\\-(\alpha R+2)(b\alpha-1)e^{-\alpha(R-a-b)}+(\alpha R+2)(a\alpha-1)e^{-\alpha(R-2a)}\end{bmatrix}$$

$$+\left(\frac{b^2\alpha^2+2b\alpha+2}{\alpha^5 R}\right)\left[-(b\alpha+1)e^{-\alpha(R+2b)}+(a\alpha+1)e^{-\alpha(R+a+b)}-(b\alpha-1)e^{-\alpha R}+(a\alpha-1)e^{-\alpha(R-a+b)}\right]$$

$$-\left(\frac{a^2\alpha^2+2a\alpha+2}{\alpha^5 R}\right)\left[-(b\alpha+1)e^{-\alpha(R+a+b)}+(a\alpha+1)e^{-\alpha(R+2a)}-(b\alpha-1)e^{-\alpha(R+a-b)}+(a\alpha-1)e^{-\alpha R}\right]$$

$$-\left(\frac{b^2\alpha^2-2b\alpha+2}{\alpha^5 R}\right)\left[-(b\alpha+1)e^{-\alpha R}+(a\alpha+1)e^{-\alpha(R+a+b)}-(b\alpha-1)e^{-\alpha(R-2b)}+(a\alpha-1)e^{-\alpha(R-a-b)}\right]$$

$$+\left(\frac{a^2\alpha^2-2a\alpha+2}{\alpha^5 R}\right)\left[-(b\alpha+1)e^{-\alpha(R-a+b)}+(a\alpha+1)e^{-\alpha R}-(b\alpha-1)e^{-\alpha(R-a-b)}+(a\alpha-1)e^{-\alpha(R-2a)}\right]$$

**A5**

These expressions have been used for the calculations of potential energies.